\documentclass[a4paper,twocolumn,accepted=2024-04-10]{quantumarticle}
\pdfoutput=1
\usepackage[utf8]{inputenc}
\usepackage[english]{babel}
\usepackage[T1]{fontenc}
\usepackage{amsmath}
\usepackage{amsfonts}
\usepackage{amssymb}
\usepackage{graphicx}
\usepackage{hyperref}
\usepackage{xcolor}
\usepackage{float}
\usepackage{physics}
\usepackage{ulem}
\usepackage[numbers]{natbib}

\newcommand{\bsy}[1]{\ensuremath{\boldsymbol{#1}}}

\begin{document}
\title{Photonic entanglement with accelerated light}

\author{R. C. Souza Pimenta }
\affiliation{Departamento de F\'{i}sica, Universidade Federal de Santa Catarina, CEP 88040-900, Florian\'{o}polis, SC, Brazil}

\author{G. H. dos Santos}
\email{fisica.gu@gmail.com}
\affiliation{Departamento de F\'{i}sica, Universidade Federal de Santa Catarina, CEP 88040-900, Florian\'{o}polis, SC, Brazil}

\author{A. B. Barreto}
\email{adriano.barreto@caxias.ifrs.edu.br}
\affiliation{Departamento de F\'{i}sica, Universidade Federal de Santa Catarina, CEP 88040-900, Florian\'{o}polis, SC, Brazil}
\affiliation{Instituto Federal de Educa\c{c}\~{a}o, Ci\^{e}ncia e Tecnologia do Rio Grande do Sul (IFRS), \textit{Campus} Caxias do Sul, RS, Brazil}

\author{L. C. C\'{e}leri}
\email{lucas@qpequi.com}
\affiliation{QPequi Group, Institute of Physics, Federal University of Goi\'{a}s, 74690-900, Goi\^{a}nia, GO, Brazil}

\author{P. H. Souto Ribeiro}
\email{p.h.s.ribeiro@ufsc.br}
\affiliation{Departamento de F\'{i}sica, Universidade Federal de Santa Catarina, CEP 88040-900, Florian\'{o}polis, SC, Brazil}

\begin{abstract}
Accelerated light has been demonstrated with laser light and diffraction. Within the diffracting field it is possible to identify a portion that carries most of the density beam energy, which propagates in a curved trajectory as it would have been accelerated by a gravitational field for instance. Here, we analyze the effects of this kind of acceleration over the entanglement between twin beams produced in spontaneous parametric down-conversion. Our results show that this type of acceleration does not affect entanglement significantly. The optical scheme introduced can be useful in the understanding of processes in the boundary between gravitation and quantum physics.

\end{abstract}
%
%
\maketitle 
\section{Introduction\label{secIntroduction}}

Optical systems have been of primary relevance in the discovery and understanding of quantum entanglement and its applications to quantum information, from Bell's inequality~\cite{Clauser72,Aspect82,Zeilinger98} to applied quantum technology~\cite{JPan1, zeilingertele97,Zia23}. An important part of these contributions has become possible thanks to the nonlinear optical process known as  spontaneous parametric down-conversion (SPDC)~\cite{Burnham70}. In this process, one pump photon is converted into a pair of photons called signal and idler. They can be prepared in entangled states related to a few different degrees of freedom like polarization, time bins, orbital angular momentum and transverse spatial degrees of freedom~\cite{Tittel01}. Due to energy and momentum conservation in the down-conversion process, signal and idler photons are generated practically at the same position inside a nonlinear optical medium and are emitted in anti-correlated directions, which results in a naturally entangled state for the transverse spatial degrees of freedom~\cite{Walborn10}.

One way of demonstrating that signal and idler photons are non-separable with respect to the transverse spatial position and momentum is through the violation of inequalities that are based on the variances of conditional measurements of these quantities~\cite{dgcz2000,mgvt2002}. This kind of test is sufficient but not necessary for proving entanglement, thus representing an entanglement witness. There have been several experimental realizations of such tests demonstrating that signal and idler photons are actually non-separable~\cite{Boyd04}. Here we are interested in the same kind of non-separability test, but when light is accelerated. Such a scenario is interesting in many contexts, but especially in order to simulate the effects of the equivalence principle on light or even flat spacetime phenomena, like the Unruh effect~\cite{Martinez-14}. 

Experimentally, we create an Airy beam using a spatial light modulator that applies a specific diffraction mask to a Gaussian beam, converting it into an Airy beam~\cite{Siviloglou:07,Efremidis:19}. The Airy beams are called accelerated, even though they do not violate Ehrenfest's theorem or the conservation of transverse electromagnetic momentum. The Airy-acceleration effect has to do with the intensity structure, as explained in Ref. \cite{Efremidis:19}.
Therefore, we experimentally test an entanglement witness for transverse spatial entanglement, when the idler photon is Airy-accelerated. We show that signal and idler photons are still entangled. We provide a quantum optical theoretical description and present a discussion on the possible utilization of this experimental scheme to study gravitation effects on quantum entanglement.
 
\section{Theory}
\label{theo}

We would like to test the effect of Airy-acceleration on entanglement in a configuration as simple as possible. Let us consider a down-conversion photon pair source and the down-converted beams are called signal and idler arbitrarily. The signal beam propagates freely to a single photon avalanche diode (SPAD), while the idler one is Airy-accelerated by means of a spatial light modulator (SLM) diffraction mask and is also detected by a SPAD. Signal and idler photons are detected in coincidence and the conditional transverse spatial distributions of position and momentum are measured by means of imaging and Fourier transforming optical schemes. From these measurements, the conditional variances are obtained and used to perform non-separability tests.

The results are compared to measurements previously performed with both signal and idler beams measured after free propagation, demonstrating that the down-conversion source produces entangled photon pairs. In our experiment, we employ diffraction by a phase-mask that is able to convert plane waves into Airy-accelerated. In practice, Gaussian beams are converted into approximate Airy beams inside an envelope Gaussian function. Despite the approximation, these beams reproduce all relevant properties of the Airy beams within a finite propagation range. It is of particular importance the fact that they propagate along a curved trajectory just like the ideal Airy beams.

\subsection{Diffraction theory}

We follow the same lines of the calculations presented in Ref.~\cite{Fonseca99}, which derives the coincidence counting rate between signal and idler detectors as a function of their transverse spatial positions. 
In Ref.~\cite{Fonseca99}, it is considered that signal and idler beams are both modulated in phase and amplitude. However, our approach is simpler and modulates only one among the two beams. 

Let us start by writing the electric field operator for a monochromatic field propagating along the $z$ axis in the plane $z=0$
\begin{equation}
\label{EF}
\bsy{E^{+}(\bsy{\rho},\mbox{0}) = \int dq \,\, a(q) \, e^{iq.\rho}},
\end{equation}
where $\bsy\rho$ is a position vector in a plane orthogonal to the $z$ axis, $\bsy q$ is the transverse component of the wavevector $\bsy k$, and $\bsy a$ is the annihilation operator.

After free propagation from the crystal source located at $z = 0$ to the plane $z$, the electric field operator
for the signal field can be written as
\begin{equation}
\label{EFF}
\bsy E_s^+(\bsy \rho_s,\mbox{z}) = \int d\bsy q_s \,\, \bsy a_s(q) \, \exp[i\left(\bsy q_s.\bsy \rho_s- \frac{q_s^2}{2k_s}z\right)],
\end{equation}
where $k$ is the wavenumber and $q = |\bsy q|$.

The idler field propagates a distance $z_A$ from the crystal to the SLM diffracting mask and then a distance $z - z_A$ from the SLM to the detector. The electric field operator for the idler in this case is given by~\cite{monken:98}
\begin{eqnarray}
\label{EEFF}
&\bsy E_i^+(\bsy \rho_i,\mbox{z}) = \int d\bsy q_i \int d\bsy q'_i \,\, \bsy a_i(\bsy q_i) \, \bsy T(\bsy q_i - \bsy q'_i)& \nonumber \\
&\times  \mbox{exp}\left[i\left(\bsy{q}_i \cdot\bsy{\rho}_s - \frac{\bsy q_i^2}{2k_i}(z - z_A) - \frac{\bsy q_i^{'2}}{2k_i}z_A\right)\right], &
\end{eqnarray}
where $\bsy T (\bsy q)$ is the transfer function of the SLM modulation function. In this experiment the SLM is programmed with the Fourier transfer of the Airy function, therefore in the above equation we have the Airy function written in terms of the transverse momentum variables. $z_a$ is the distance between the crystal and the SLM, and $(z-z_A)$ is the distance between the SLM and the detection plane.

Equations~\ref{EFF} and~\ref{EEFF} provide the electric field operators for signal and idler fields at the detection plane, written in terms of the fields in the source. Therefore, in order to obtain the coincidence counting rate at signal and idler detection planes, we just apply these operators on the state of the field at the source plane and take the square modulus. For simplicity, we will consider that the state at the source is an ideal EPR state given by
\begin{equation}
|\Psi^{EPR}\rangle = \int d\bsy q_1 \int d\bsy q_2 \, \delta(\bsy q_1 + \bsy q_2) \ket{\bsy q_1} \ket{\bsy q_2},
\label{EPR}
\end{equation}
with $\bsy q_{1(2)}$ being the transverse momentum vectors for signal and idler, and $\ket{\bsy q_m}_m$, with $m = 1(2)$, represents a single photon state populating a mode with transverse wave vector component $\bsy q$.

The coincidence counting rate is given by
\begin{equation}
C(\bsy \rho_i,\bsy \rho_s)  = |c(\bsy \rho_i,\bsy \rho_s)|^2,
\label{G2}
\end{equation}
where $\bsy \rho_i$ and $\bsy \rho_s$ are position vectors at idler and signal detection planes respectively, and
\begin{eqnarray}
c(\bsy \rho_1,\bsy \rho_2) = \langle 0,0| \bsy E_i^+ \bsy E_s^+ |\Psi^{EPR}\rangle,
\end{eqnarray}
where $\langle 0,0|$ is the vacuum state. Calculation of this quantity gives
\begin{eqnarray}
\label{psi}
c(\bsy \rho_1,\bsy \rho_2) &=& \int \dd\bsy\xi \, \bsy T (\bsy\xi) \, \, \mbox{exp}\left(i k \frac{\xi^2}{4z_a}\right)  \nonumber \\
&\times& \mbox{exp}\left(i k \frac{|\frac{1}{2}(\bsy\rho_1 - \bsy\rho_2) - \bsy\xi|^2}{z_d - z_a}\right)  \nonumber \\ 
&=& \bsy{Ai} (\bsy\rho_1 - \bsy\rho_2),
\end{eqnarray}
where $\bsy T (\bsy\xi)$ is the Fourier transform of the Airy function of the first kind $\bsy{Ai} (\bsy q)$. In words, the calculation in Eq. \eqref{psi} describes the Fresnel transform of the field $ \bsy T (\bsy\xi) \mbox{exp}\left(i k \frac{\xi^2}{4z_a}\right)$, which propagates it a distance $z - z_A$ in the paraxial approximation. If this distance is large enough, the Fresnel transform is equivalent to the Fourier transform. 

The coincidence counting rate is given by
\begin{equation}
C(\rho_1,\rho_2) = |\bsy{Ai} (\bsy{\rho}_1 - \bsy{\rho}_2)|^2.
\label{airy}
\end{equation}

From the calculation above we can infer the two-photon state at the detection planes.
\begin{equation}
|\Psi\rangle = \int \dd\bsy{\rho}_1 \int \dd\bsy{\rho}_2 \, \bsy{Ai}(\bsy{\rho}_1 - \bsy{\rho}_2) |\bsy \rho_1 \rangle_1 |\bsy \rho_2\rangle_2.
\label{airystate}
\end{equation}

This state is still pure, even though the generation of the Airy-acceleration in the idler beam produces losses. This is due to the fact that the two-photon correlation function considers only the cases where coincidence counts are obtained by the detectors. In other words, this is the post-selected two-photon state. The entanglement depends on the separability of the Airy function. 

\subsection{MGVT Separability criterion}

The Mancini-Giovannetti-Vitali-Tombesi (MGVT) separability witness was introduced by Mancini et al. in Ref. \cite{mgvt2002}. They show that a bipartite state
concerning continuous variables is non separable if
\begin{equation}
\label{mgvt1}
\Delta(p_1 + p_2) \Delta(x_1 - x_2) < |[x,p]|^2,
\end{equation}
where $\Delta$ is the variance, $x_1,p_1$ and $x_2,p_2$ concerns subsystems 1 and 2 respectively and $[x,p]$ is the commutator between $x$ and $p$. 

The derivation of this inequality can be readily done by calculating the above variances for a general separable state
\begin{equation}
\label{mgvt2}
\rho = \sum_i p_i \rho_1 \otimes \rho_2,
\end{equation}
and using the Schwarz inequality
\begin{equation}
\label{mgvt3}
\left(\sum_i p_i\right) \left(\sum_i p_i \langle v \rangle_i^2\right) \geq \left(\sum_i p_i |\langle v_i \rangle| \right)^2 ,
\end{equation}
where $v$ is an operator. The inequality obtained must be obeyed by separable states: 
\begin{equation}
\label{mgvt4}
\Delta(p_1 + p_2) \Delta(x_1 - x_2) \geq |[x,p]|^2.
\end{equation}

Therefore, violation of inequality \eqref{mgvt4} implies that the state is non separable. Notice that there are other
witnesses, like the one derived by Duan et al. \cite{dgcz2000}. We have chosen MVGT criterion for convenience and it helps showing that the two-photon state produced in spontaneous parametric down conversion is still non separable, even when the idler (or signal) beam is accelerated.

\section{Experiment}

\begin{figure}[ht]
    \centering
    \includegraphics[width = 0.4\textwidth]{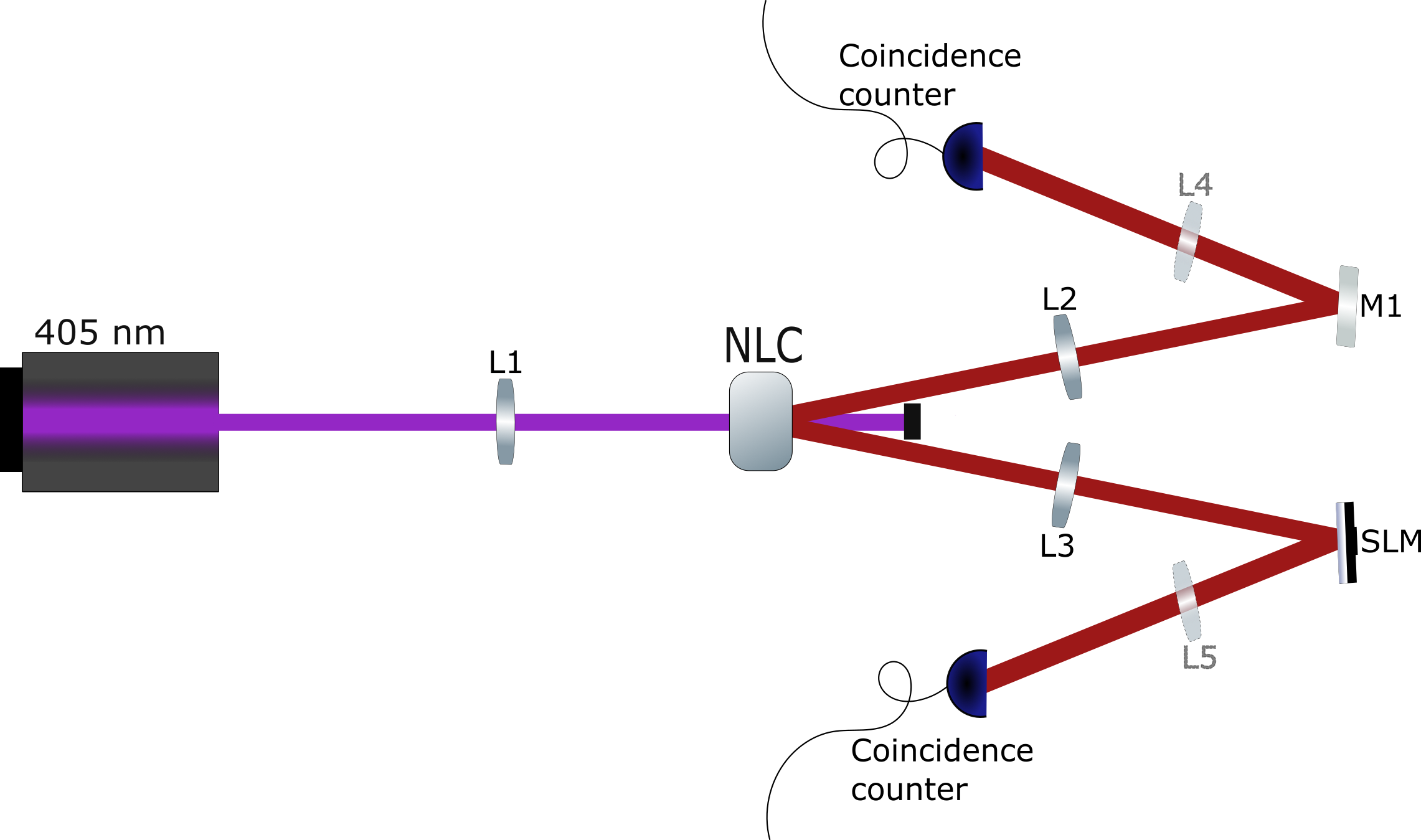}
    \caption{Experimental setup for measuring coincidences produced by a pair of photons from SPDC. ``$@$405'' is a blue laser at 405 nm, ``NLC'' is a nonlinear crystal, ``L1'', ``L2'', and ``L3'' are lenses that are always on the setup. With L1 we focus the pump ahead of NLC to increase the spatial correlation.  ``M1'' is a mirror, ``SLM'' is the Spatial Light Modulator used to apply the phase mask. L2 and L3 are used to make the image of the NLC on both SLM and mirror. ``L4'' and ``L5'' are lenses placed only when measuring position distributions.}
    \label{setup}
\end{figure}

Figure~\ref{setup} shows the sketch of the experimental setup. A diode laser oscillating at 405~nm is used to pump a BBO nonlinear crystal and produce spontaneous parametric down conversion.  Degenerate twin beams at 810~nm are selected by means of interference filters placed in front of the detectors. The signal beam is directed to the single-photon counting module (SPCM) through exchangeable sets of lenses. One set of lenses image the source plane onto the detection plane, while the other set performs the optical Fourier transform. With these two sets of lenses it is possible to measure position and momentum of signal photons.  The idler photon is also passed through two exchangeable sets of lenses in order to measure position and momentum. However, it is also reflected by a spatial light modulator (SLM), which applies a phase mask. The spatial modulation is designed to couple the idler photon onto an Airy-accelerated beam. Signal and idler beams are both detected with SPCMs and coincidence counting is performed during a measurement time of 10 seconds in each run. 

Even though we have used a phase-only modulator, amplitude modulation is also performed by using a diffraction mask combined with the modulation function and detecting only the first order diffraction beam. Phase-only modulation schemes with entangled photon pairs have been investigated in the context of the so called ghost images \cite{abouraddy}.

We start by performing the measurements of the conditional variances for free propagating signal and idler beams.  In order to measure the conditional distribution of birth-positions of photons in the source, a pair of lenses is used to image the crystal face onto the detection plane for both signal and idler beams with a magnification factor of about 3.  While the signal detector is kept fixed near the coincidence peak position, the idler detector is scanned along the vertical axis with respect to the optical table. The measurement of the conditional momentum distribution follows the same protocol, but replacing the imaging lenses with Fourier transforming ones. This operation maps the momentum distribution of signal and idler photons at the crystal face onto the detection plane. During all measurements, we also register the single photon counting rates for both detectors. 

The second step is repeating all the procedure for measuring the conditional distributions of position and momentum when the idler beam is Airy-accelerated.

We change a parameter of the SLM diffraction mask, so that the transverse displacement of the Airy beam peak is changed. This change can be interpreted in two equivalent ways: $i$) the acceleration increases, thus increasing the curvature of the trajectory; or  $ii$) the propagation distance is increased and the transverse deflection increases. In the following, we will treat this parameter as the propagation distance $Z$, corresponding to the case $ii$). 

The third step is to measure the conditional variances for signal and idler beams propagated to a plane far from the crystal. In the previous measurement, we have performed position and momentum measurements with respect to the plane of the crystal face.

\section{Results}

\begin{figure}[h]
    \centering
    \includegraphics[width = 0.4\textwidth]{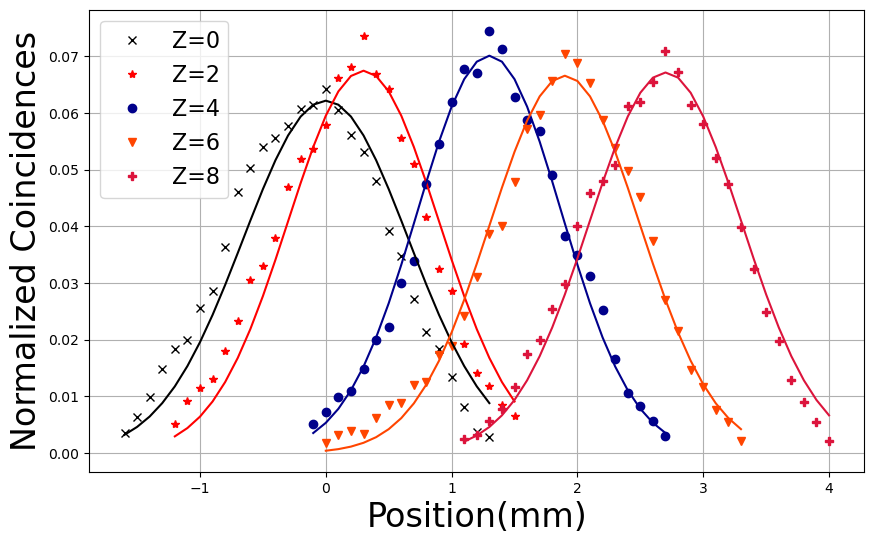}
    \caption{Plot of normalized coincidences as a function of the idler detector vertical displacement and varying the propagation distance ($Z$). The solid lines are fittings to Gaussian functions. See details in the main text.}
    \label{plot1}
\end{figure}

Figure~\ref{plot1} contains a plot of normalized coincidence counts describing the conditional momentum distributions for different values of the parameter $Z$ applied to the SLM that generates the Airy-accelerated idler beam. One can see that the coincidence peaks get displaced when $Z$ increases, which warrants the Airy-like modulation on the input idler profile.
The coincidence profiles are not exactly like Airy functions, as predicted by Eq. \eqref{airy}. This is mainly due to two technical issues. First, the SPDC source does not produce a perfect EPR state with a delta function-like correlation, and therefore, the Airy function is convoluted with a Gaussian, which better describes the SPDC state. Second, the measurement resolution of both signal and idler detectors is finite, and this smoothes the oscillation fringes of the Airy function. This loss of visibility in the Airy pattern does not invalidate the entanglement witness, because it can only increase the conditional position and momentum variances. 
In this way, we could have a false NON-violation of the separability inequality, but if a violations is observed the state must be entangled.

We have measured the product of variances for free propagating signal and idler beams in order to verify the entangled state produced by the SPDC source. The resulting product of variances $\Delta P\cdot\Delta X = 0.090 \pm 0.001$. According to MGVT criterion the inequality in Eq. \eqref{mgvt4} is violated when the product of variances is smaller than $|[x,p]|^2 = 1$ using dimensionless variables. Therefore $\Delta P\cdot\Delta X < 1$ implies in the violation. For estimating the error bars, we take the usual procedure of considering that both single photon and coincidence counting statistics are Poissonian. This provides uncertainties for the coincidence counts that are given by the square root of the measured values. The uncertainty of the variances comes from the propagation of the counting uncertainties to their widths.
\begin{table}[H]
\centering
\caption{Product of variances}
\label{desig}
\begin{tabular}{cc}
\hline
Z & $\Delta P\cdot\Delta X$ \\ \hline
0 & 0.24 $\pm$ 0.01                  \\
2 & 0.22 $\pm$ 0.01                  \\
4 & 0.21 $\pm$ 0.01                  \\
6 & 0.25 $\pm$ 0.01                  \\
8 & 0.22 $\pm$ 0.01\\
\hline
\end{tabular}
\end{table}

After testing the non separability of free propagating signal and idler beams, we modulate the idler beam with the SLM so that it is changed into an Airy-accelerated beam and perform the measurements for evaluating the entanglement witness again. The results are shown in Table~\ref{desig} for different values of the parameter propagation distance Z. We note that even though the product $\Delta P \cdot \Delta X$ is larger than that for free propagation, the inequality is still violated. The increment in $\Delta P$ is expected due to the enlargement of the idler transverse profile that the SLM modulation produces. However, the entanglement can always be witnessed. We also observe that the variation of the propagation distance Z does not affect the conditional far field variance significantly. To understand that, we recall that Airy wave packets are nonspreading \cite{Berry:79,Lung:08}. 
\begin{figure}[h]
    \centering
    \includegraphics[width = 0.4\textwidth]{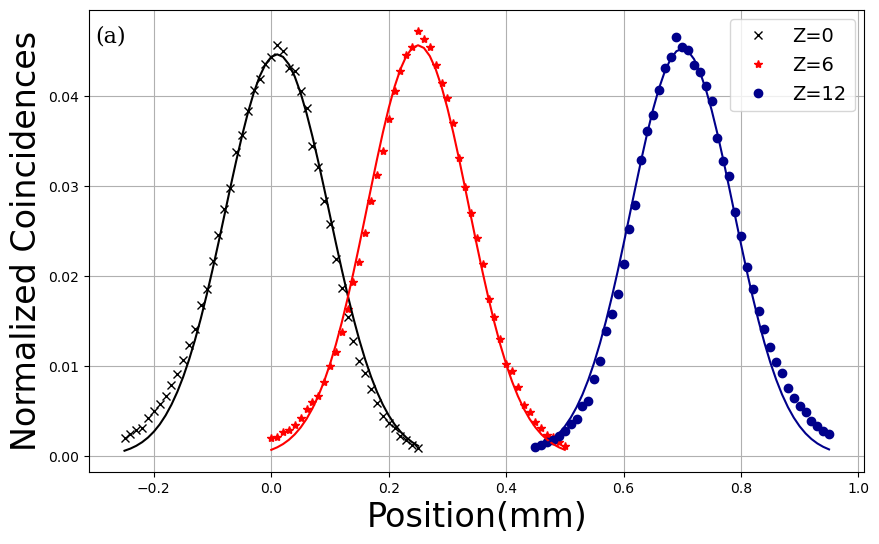}
    \includegraphics[width = 0.4\textwidth]{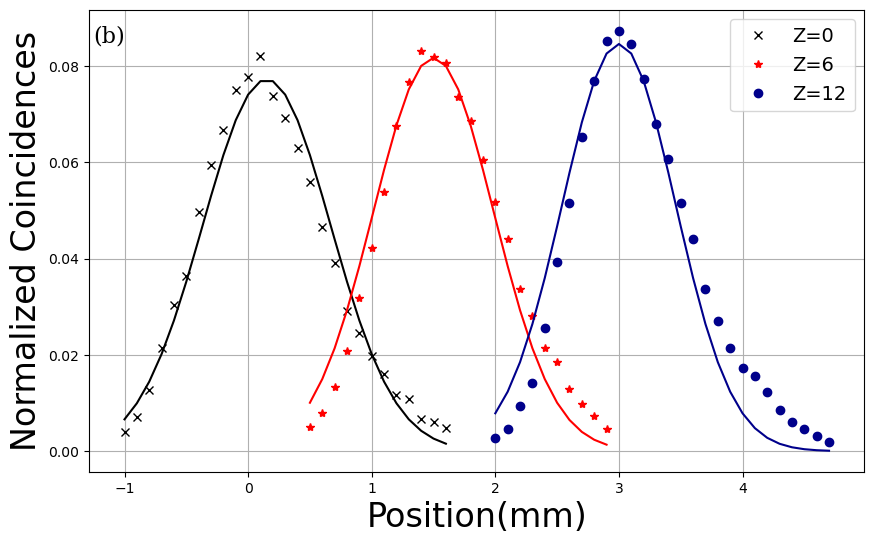}
    \caption{Plot of normalized coincidences as a function of the idler detector vertical displacement and varying the propagation distance (Z). The solid lines are fittings to Gaussian functions. (a) Plots for position-position correlation measurements. (b) Plots for momentum-momentum correlation measurements. See details in the main text.}
    \label{plot2}
\end{figure}

We also performed a measurement for the propagated field, while the previous measurement concerned the SPDC crystal facet. We make the image and the Optical Fourier Transform of a plane situated 10~cm after the SLM for the idler and 10~cm after mirror M1 for the signal beam. The results show that the state remains non-separable. The results are shown in Fig. \ref{plot2}. The upper plot shows the position correlations and the lower one shows the momentum anti-correlations. In both graphs the distance $Z$ have been varied, showing the expected transverse displacement signature of Airy-accelerated beams. Table \ref{desigplan10} shows the product variances demonstrating that the nonseparability inequality is violated.
\begin{table}[H]
\centering
\caption{Product of variances}
\label{desigplan10}
\begin{tabular}{cc}
\hline
Z & $\Delta P\cdot\Delta X$ \\ \hline
0 & 0.39 $\pm$ 0.02                  \\
6 & 0.33 $\pm$ 0.01                 \\
12 & 0.35 $\pm$ 0.01  \\
\hline
\end{tabular}
\end{table}
\section{Interpretation in terms of gravitational acceleration}

We take two photons entangled in their transverse spatial degrees of freedom and couple one of them into an Airy-accelerated mode and analyze the effect over the entanglement between them. We calculate the coincidence counting rate distribution for this case and show that it follows an Airy function. We find that the calculated final state is pure, because it describes the post-selected coincidences. The calculation shows that the post-selected state is entangled as long as the Airy function is non-separable for signal and idler coordinates. We perform an experiment demonstrating that our SPDC source produces entangled photon pairs, as expected. We show the two-photon state is also non-separable when the idler photon populates an Airy-accelerated mode. This is demonstrated with measurements taken from the SPDC crystal face and also from a distant plane after propagation of signal and idler photons. 

Berry et al.~\cite{Berry:79} studied nonspreading wavepackets in the form of Airy wavepackets. They state that the acceleration of the Airy packet is related to the curvature of the caustic of the family of world lines in spacetime. It does not violate the Ehrenfest's theorem because the acceleration is related to the intensity or probability distribution and not to the field amplitude or wavefucntion. They also emphazise that this is unique for Airy-packets. Inspired in this work, we interpret our
experiment as the idler photon prepared in an Airy-accelerated wavepacket represents the effect of a gravitational force on a rigid particle. Rigid, because the wavepacket is nonspreading. Therefore, our results might help understanding the effects of propagation in a curved spacetime over quantum entanglement between two particles and may inspire further experiments following the same reasoning. Given that experiments of this type in real curved spacetime is still very challenging, we believe that our approach can be helpful. 

It is worth mentioning that a recent work analyzed the effect of pumping a SPDC crystal
with an Airy beam laser~\cite{Lib:20b}. They measure coincidence counting rates that follow an Airy function distribution. However, this work is different from ours because the Airy-like correlation function is a build-in characteristic imprinted by the pump angular spectrum. They transfer the Airy-beam angular spectrum from the pump to the two-photon wave function in the same fashion as demonstrated in Ref.~\cite{monken:98}. In our experiment, the Airy beam shape is imprinted only in the idler beam in an asymmetric way. This difference is crucial to evaluate the effect of Airy-acceleration over the entanglement. The same difference also allows the interpretation in terms of gravitational acceleration. 

\section{Conclusion}

In conclusion, we analyzed the effect of Airy-acceleration on one beam of an entangled photon pair. We presented theoretical and experimental results showing that if one photon of the pair is accelerated, while the other is not, the non separability is preserved. We interpret the results in terms of the optical Airy beam acceleration and we also present an interpretation comparing our scheme to a massive particle accelerated by a gravitational force. We suggest that our approach can be useful to advance the understanding of gravitation effects over quantum entangled systems.

\begin{acknowledgments}
Coordena\c c\~{a}o de Aperfei\c coamento de Pessoal de N\'\i vel Superior (CAPES DOI 501100002322), 
Funda\c c\~{a}o de Amparo \`{a} Pesquisa do Estado de Santa Catarina (FAPESC - DOI 501100005667),
Conselho Nacional de Desenvolvimento Cient\'{\i}fico e Tecnol\'ogico (CNPq - DOI 501100003593), 
Instituto Nacional de Ci\^encia e Tecnologia de Informa\c c\~ao Qu\^antica (INCT-IQ 465469/2014-0).
\end{acknowledgments}
\newpage
\bibliographystyle{quantum.bst}


\end{document}